\documentclass[prd,twocolumn,showpacs,amsmath,amssymb]{revtex4}
\usepackage{graphicx}% Include figure files
\usepackage{dcolumn}% Align table columns on decimal point
\usepackage{bm}% bold math

\newcommand{\BR}{{\cal B}}

\newcommand{\psip}{\psi^{\prime}}

\newcommand{\psipp}{\psi^{\prime \prime}}

\newcommand{\jpsi}{J/\psi}
\newcommand{\ccbar}{c\overline{c}}
\newcommand{\DDbar}{D\overline{D}}

\newcommand{\EE}{e^+e^-}
\newcommand{\PP}{\pi^+\pi^-}

\newcommand{\kskl}{K^0_SK^0_L}

\newcommand{\jpsipp}{J/\psi \pi^+\pi^-}

\newcommand{\ra}{\rightarrow}
\newcommand{\jpsito}{J/\psi \rightarrow }

\newcommand{\rhopi}{\rho\pi}

\newcommand{\beq}{\begin{equation}}
\newcommand{\eeq}{\end{equation}}
\newcommand{\beqn}{\begin{eqnarray}}
\newcommand{\eeqn}{\end{eqnarray}}
\newcommand{\beqns}{\begin{eqnarray*}}
\newcommand{\eeqns}{\end{eqnarray*}}
\newcommand{\bfg}{\begin{figure}}
\newcommand{\efg}{\end{figure}}
\newcommand{\bitm}{\begin{itemize}}
\newcommand{\eitm}{\end{itemize}}
\newcommand{\bnum}{\begin{enumerate}}
\newcommand{\enum}{\end{enumerate}}
\newcommand{\btbl}{\begin{table}}
\newcommand{\etbl}{\end{table}}
\newcommand{\btbu}{\begin{tabular}}
\newcommand{\etbu}{\end{tabular}}
\begin{document}

\title{Measuring $\psipp \ra \kskl$ as a test of the $S$- and $D$-wave
mixing of charmonia}
\date{\today}

\author{P.~Wang}
 \email{wangp@mail.ihep.ac.cn}
\author{X.~H.~Mo}
\author{C.~Z.~Yuan}
\affiliation{Institute of High Energy Physics, CAS, Beijing 100039, China}
%%%\author{TPG}

\begin{abstract}
Adding to the long standing ``$\rhopi$ puzzle'' in $\psip$ and $\jpsi$ decays, 
recently BEijing Spectrometer (BES) reported $\BR(\psip\ra\kskl)$ which is 
enhanced relative to the pQCD ``$12\%$ rule'' expectation from
$\BR(\jpsito \kskl)$. If the enhancement is due to the mixing of
the $S$- and $D$-wave charmonium states as in the $\rhopi$ case, the
newly measured $\BR(\psip\ra\kskl)$ gives a constraint on
$\BR(\psipp\ra \kskl)$. It serves as a good test for the scenario
of the $S$- and $D$-wave mixing in the $\psip$ and $\psipp$. \vspace{1pc}
\end{abstract}


\maketitle

\section{Introduction}

From the perturbative QCD (pQCD), it is expected that both $\jpsi$
and $\psi(3686)$ (shortened as $\psip$) decaying into light hadrons
are dominated by the annihilation of $\ccbar$ into three gluons,
with widths proportional to the square of the wave function
at the origin~\cite{appelquist}. This yields the pQCD ``12\% rule'', that is
\begin{eqnarray}
Q_h &=&\frac{{\cal B}_{\psip \ra h}}{{\cal B}_{\jpsi \ra h}}
=\frac{{\cal B}_{\psip \ra \EE}}{{\cal B}_{\jpsi \ra \EE}} \approx
 12\%~~.
\label{qcdrule}
\end{eqnarray}

The violation of the above rule was first observed in $\rhopi$ and 
$K^{*+}K^-+c.c.$ modes by Mark II~\cite{mk2}, since then 
BES has measured many two-body decay modes of $\psip$, among which some
obey the 12\% rule while others violate it~\cite{guyf}. There
have been many theoretical efforts trying to solve the
puzzle~\cite{puzzletheory}, however, none explains all the
existing experimental data satisfactorily and naturally~\cite{revpuz}.

A most recent explanation of the ``$\rhopi$ puzzle'' using the
$S$- and $D$-wave charmonia mixing was proposed by Rosner~\cite{rosnersd}.
In this scheme, the mixing of  $\psi (2^3 S_1)$ state
and $\psi (1^3 D_1)$ is in such a way which leads to almost complete
cancellation of the decay amplitude of $\psip \rightarrow \rhopi$,
and the missing $\rhopi$ decay mode of $\psip$ shows up instead as
enhanced decay mode of $\psi(3770)$ (shortened as $\psipp$). A study
on the measurement of $\psipp \ra \rhopi$ in $\EE$ experiments shows that
with the decay rate predicted by the $S$- and $D$-wave mixing, the
destructive interference between the three-gluon decay amplitude of the
$\psipp$ resonance and the continuum one-photon amplitude
leads to a very small cross section~\cite{wympspp}, 
which is in agreement with the unpublished upper limit of the $\rhopi$ cross 
section at the $\psipp$ peak by Mark III~\cite{mk3}. Although
this needs to be further tested by high luminosity experiment
operating at the $\psipp$ mass energy, such as CLEO-c, it already
showed that ${\cal B}(\psipp \rightarrow \rhopi)$ is most probably
at the order of $10^{-4}$, in agreement with the prediction of the
$S$- and $D$-wave mixing scheme.

If the $S$- and $D$-wave mixing is the key for solving the $\rhopi$
puzzle, it applies to other decay modes as well, such as
pseudoscalar pseudoscalar (PP) mode like $\kskl$.
Recently, BES collaboration reported the $\kskl$ branching ratios
of $\jpsi$ and $\psip$ decays~\cite{besjpsi,bespsip}:
\beqn
\BR(\jpsi \ra\kskl)=(1.82\pm0.04\pm0.13)\times10^{-4}~~,
\nonumber \\
\BR(\psip \ra \kskl)=(5.24\pm0.47\pm0.48)\times10^{-5}~~.
\label{brkskl}
\eeqn
These results yield $Q_{\kskl} =(28.8 \pm 3.7)\%$, which is enhanced
relative to the 12\% rule by more than 4$\sigma$.
In this paper, the $\psip \ra \kskl$ enhancement is explained in the 
$S$- and $D$-wave charmonia mixing scheme, and the $\psipp \rightarrow \kskl$ 
decay rate is estimated with the inputs $\BR(\jpsi \ra \kskl)$ and $\BR(\psip
\ra \kskl)$. In following sections, the mixing scheme is introduced briefly, 
then the branching ratio of $\psipp \ra \kskl$ is calculated with the
measured $\EE$ and $\kskl$ decay rates of $\jpsi$ and $\psip$,
assuming the mixing of $S$- and $D$-wave. Finally the experiment
search for $\psipp \rightarrow \kskl$ is proposed.

\section{$S$- and $D$-wave Mixing scheme}

To explain the measured $\Gamma_{ee}$ of $\psipp$, it is
suggested~\cite{eichten2,kuang,rosner} that the mass eigenstates $\psip$ and
$\psipp$ are the mixtures of the $S$- and $D$-wave
of charmonia, namely $\psi (2^3 S_1)$ and $\psi (1^3 D_1)$ states.
In this scheme,
\beq
\begin{array}{l}
 |\psip\rangle = | 2^3 S_1 \rangle \cos \theta
                  - | 1^3 D_1 \rangle \sin \theta~, \\
 |\psipp\rangle = | 2^3 S_1 \rangle \sin \theta
                  + | 1^3 D_1 \rangle \cos \theta~,
\end{array}
\label{sdmix}
\eeq
where $\theta$ is the mixing angle between pure $\psi(2^3 S_1)$
and $\psi(1^3D_1)$ states and is fitted from the leptonic widths of $\psipp$
and $\psip$ to be either $(-27 \pm 2)^{\circ}$ or
$(12 \pm 2)^{\circ}$~\cite{rosnersd}.
The latter value of $\theta$ is consistent with the coupled channel
estimates~\cite{eichten2,heikkila} and with the ratio of
$\psip$ and $\psipp$ partial widths to $\jpsipp$~\cite{kuang}.
Hereafter, the discussions in this paper are solely for the
mixing angle $\theta =12^{\circ}$.

As in the discussion of Ref.~\cite{rosnersd}, since both hadronic
and leptonic decay rates are proportional to the square of the wave
function at the origin $|\Psi(0)|^2$, it is expected that if $\psip$ is
a pure $\psi(2^3S_1)$ state, then for any hadronic final states $f$,
\beq
\Gamma(\psip \ra f)=\Gamma(\jpsi \ra f) \frac{\Gamma(\psip \ra \EE)}
{\Gamma(\jpsi \ra \EE)}~.
\label{asu}
\eeq

The electronic partial width of $\jpsi$ is expressed in potential
model by~\cite{novikov}
\beq
\Gamma(\jpsi \ra \EE)=\frac{4\alpha^2 e^2_c}{M_{\jpsi}^2}
\left|R_{1S}(0)\right|^2,
\label{jee}
\eeq
with $\alpha$ the QED fine structure constant, $e_c=2/3$, $M_{\jpsi}$ the 
$\jpsi$ mass and $R_{1S}(0)$ the radial $1^3S_1$ wave function at the origin.

$\psip$ is not a pure $\psi(2^3S_1)$ state, its electronic partial width is
expressed as~\cite{rosnersd}
\beqn
\Gamma(\psip \ra \EE) & =&
\frac{4\alpha^2 e^2_c}{M_{\psip}^2}
  \label{sipee} \\
&\times& \left|\cos \theta R_{2S}(0) - \frac{5}{2\sqrt{2}m_c^2}
\sin \theta R^{\prime\prime}_{1D}(0) \right|^2~,  \nonumber
 \eeqn
with $M_{\psip}$ the $\psip$ mass,
$R_{2S}(0)$ the radial $2^3S_1$ wave function at the origin and
$R^{\prime\prime}_{1D}(0)$ the second derivative of the radial
$1^3D_1$ wave function at the origin.

If Eq.~(\ref{asu}) holds for a pure $2^3S_1$ state, $\psipp \ra f$,  
$\psip \ra f$ and $\jpsi \ra f$ partial widths are to be
\beqn
\Gamma(\psipp \ra f) & = & \frac{C_f}{M_{\psipp}^2} \left|
\sin \theta R_{2S}(0) + \eta \cos \theta \right|^2,
     \nonumber \\
\Gamma(\psip \ra f) & = & \frac{C_f}{M_{\psip}^2}
\left| \cos \theta R_{2S}(0) - \eta \sin \theta \right|^2,
       \nonumber \\
\Gamma(\jpsi \ra f) & = & \frac{C_f}{M_{\jpsi}^2} \left|R_{1S}(0) \right|^2~,
\label{tof}
\eeqn
where $C_f$ is a common factor for the final state $f$, $M_{\psipp}$ 
the $\psipp$ mass, and $\eta=|\eta| e^{i\phi}$ is a complex parameter 
with $\phi$ being the relative phase between $\langle f|1^3D_1 \rangle$ and
$\langle f|2^3S_1 \rangle$.

\section{Upper and lower bounds of $\BR(\psipp\ra \kskl)$ }\label{bornsct}

With Eqs.~(\ref{jee}, \ref{sipee}, \ref{tof}), the following two
equations are derived : \beqn \frac{\Gamma(\psip \ra
f)}{\Gamma(\jpsi \ra f)} & =& \frac{\Gamma(\psip \ra
\EE)}{\Gamma(\jpsi \ra \EE)}
  \label{fit} \\
 &\times& \left|\frac{\cos \theta R_{2S}(0) - \eta \sin\theta}
    {\cos\theta R_{2S}(0) - {\displaystyle \frac{5}{2\sqrt{2}m_c^2}}
    \sin\theta R^{\prime\prime}_{1D}(0)}  \right|^2, \nonumber
\eeqn
and
\beq
\frac{\Gamma(\psipp \ra f)}{\Gamma(\psip \ra f)} =
\frac{M_{\psip}^2}{M_{\psipp}^2}
\left|\frac{\sin \theta R_{2S}(0) + \eta \cos\theta}
    {\cos\theta R_{2S}(0) - \eta\sin\theta} \right|^2.
\label{resu}
\eeq
It is easy to see that if $\theta=0$, i.e. $\psip$ were a pure 
$\psi(2^3S_1)$ state, Eq.~(\ref{fit}) becomes Eq.~(\ref{asu}).

In the following, the discussion focuses on $f=\kskl$ final state.
The partial widths of $\psip$ and $\jpsi$ to $\EE$~\cite{pdg}
and $\kskl$~\cite{besjpsi,bespsip} are all measured by experiments;
$R_{2S}(0)=0.734~\hbox{GeV}^{3/2}$ and
$5R^{\prime\prime}_{1D}(0)/(2\sqrt{2}m_c^2) =
0.095~\hbox{GeV}^{3/2}$ are given in Ref.~\cite{rosnersd},
so for the final state $\kskl$, Eq.~(\ref{fit}) has only one unknown
variable  $\eta$. Since $\eta$ is complex, for any given phase, its
module can be determined. Then with $\eta$ substituting into
Eq.~(\ref{resu}), $\Gamma(\psipp \ra \kskl)$ can be calculated.

\begin{figure}[h] \centering
\includegraphics[height=5.5cm,width=6.5cm]{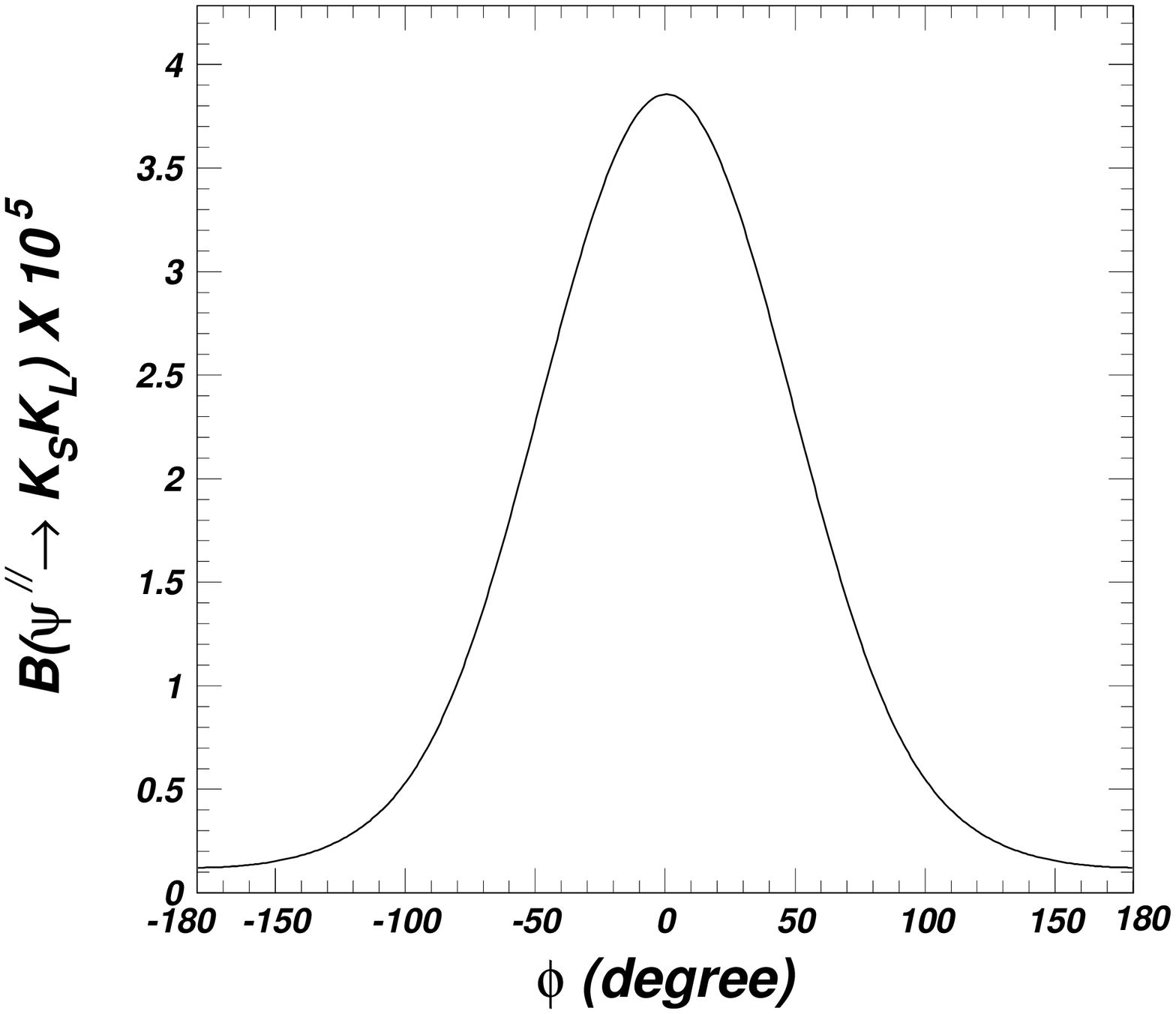}
\caption{\label{klksfig} The variation of $\BR(\psipp \ra \kskl) \times 
10^{5} $ against the phase $\phi$ (in degree).}
\end{figure}

Since the phase of $\eta$ is a free parameter, so the decay rate of 
$\psipp \ra \kskl$ is constrained in a range. According to Eqs.~(\ref{fit}) 
and (\ref{resu}), the variation of branching ratio against the phase
is shown in Fig. ~\ref{klksfig}, from which we see that
%%%%%In this way, Eqs.~(\ref{fit}) and (\ref{resu}) give 
\beq 
0.12 \pm 0.07 \le 10^{5}\times
\BR(\psipp \ra \kskl) \le 3.8 \pm 1.1~. \label{bound} 
\eeq
Here the upper bound corresponds to $\phi= 0^\circ$ 
and the lower bound to $\phi=\pm 180^\circ$. 
The uncertainties are due to the mixing angle $\theta$, 
the measurements of $\BR(\psip \ra \kskl)$ and $\BR(\jpsi \ra \kskl)$, 
with the first two dominate.

\section{Experimental test}

It it instructive to look at the range of the phase $\phi$ from other decay 
modes, such as $\rhopi$. 
The recent phenomenological estimation~\cite{largephase} gives 
the branching ratio of $\psip \ra \rhopi$ at the level of $10^{-4}$, 
which indicates the almost complete cancellation between 
$\cos \theta R_{2S}(0)$ and $\eta \sin \theta$ in Eq.~(\ref{tof}).
In another word, the small $\BR(\psip \ra \rhopi)$ means the phase 
$\phi$ of $\eta$ is around zero.
With incomplete cancellation between $\cos\theta R_{2S}(0)$
and $\eta\sin\theta$ which results in 
$\BR(\psip \ra \rhopi)=1.11 \times 10^{-4}$~\cite{largephase}, and
latest results by BES of $\BR(\jpsi \ra \rhopi) \sim 2.1\%$~\cite{jpsi3pi},
$\phi$ is constrained to be less than $11^\circ$. As a pedagogical guess, 
$\phi$ is expected to be small for other decay modes too. 
In such case, the prediction $\BR(\psipp \ra \kskl)$ would be close to 
the upper bound in Eq.~(\ref{bound}), that is
\beq 
\BR(\psipp \ra \kskl) \approx (3.8 \pm 1.1) \times 10^{-5}~. 
\label{predict} 
\eeq

Currently, BES has accumulated about 20~pb$^{-1}$ data while
CLEO-c has collected 55~pb$^{-1}$ data at $\psipp$ peak. By virtue of 
Eq.~(\ref{predict}), assuming 40\% efficiency for detecting $K^0_S \ra \PP$,
then one expects 1.7 events from BES and 4.6 events from CLEO-c.
Utilizing these samples, most probably an upper limit can be set by
BES, while the signal can be seen at CLEO-c. With the expected larger $\psipp$
data sample of several fb$^{-1}$~\cite{cleoc} in immediate future, CLEO-c 
can give a definite answer for prediction of Eq.~(\ref{predict}), 
or test the lower bound of Eq.~(\ref{bound}) in case the phase $\phi$ is
not small. 

\section{Discussion}

In the $S$- and $D$-wave mixing scheme, the observed $\psip \ra \kskl$ 
enhancement relative to the 12 \% rule implies a $\psipp \ra \kskl$
decay branching ratio at the order of $10^{-6}$ to $10^{-5}$. 
So the measurement of $\BR({\psipp \ra \kskl})$ will provide a
clear-cut test of the $S$- and $D$-wave mixing scenario.

Unlike the $\rhopi$ modes, $\kskl$ mode of the $1^{--}$ charmonium decay is 
only through strong interaction due to SU(3) symmetry~\cite{haber}. There is no complication of electromagnetic interaction and continuum one-photon 
annihilation as well as the interference between them~\cite{inter}.
So the observed $\kskl$ in $\EE$ experiment is completely from resonance
decays. 

If the $\psip$ and $\psipp$ are indeed the $S$- and $D$-wave charmonia
mixtures, not only the vector pseudoscalar~\cite{rosnersd} and the pseudoscalar pseudoscalar modes will be affected, but all the other modes in $\psip$ decays 
will be affected as well, such as vector tensor, axial-vector pseudoscalar and 
so forth. For the decay modes which have been measured
both at $\psip$ and $\jpsi$, the corresponding branching ratio at $\psipp$ 
can be evaluated under the assumption of pQCD. 
Then the measurements at $\psipp$ provide a test for the mixing scheme, 
at the same time help to reveal the charmonium decay dynamics 
and the relation between $\jpsi$ and $\psip$ decays.

The mixing scheme is a simple and natural model, %%. If it is correct, 
it will provide a new angle of purview of understanding the $\rhopi$ puzzle 
between $\jpsi$ and $\psip$ decays, and the non-$\DDbar$ decay of $\psipp$.

\section{Summary}

In this paper, the $S$- and $D$-wave mixing scheme of charmonium states is
applied on $\psip \ra \kskl$ to explain its enhancement relative to the pQCD
12\% rule, and the branching ratio of $\psipp \ra \kskl$ is predicted. 
It is suggested that with the data samples collected currently
and the larger data sample expected from CLEO-c soon, the mixing scheme 
is to be tested.

\acknowledgments
This work is supported by 100 Talents Program of the Chinese Academy of
Sciences under contract No. U-25.

\end{document}